\documentstyle[aps,twocolumn,floats,array]{revtex}

\def\(#1){(\ref{#1})}		
\def\ie{{\it i.e.\/}}

\def\kv{{\bf k}}
\def\vv{{\bf v}}

\def\rv{{\bf r}}
\def\dv{{\bf d}}

\def\g{\gamma}

\def\eps{\epsilon}
\def\G{\Gamma}
\def\sv{{\bf S}}
\def\GS{G_{\bf S}}
\def\GSk{G_{\bf S}^{\bf k}}
\def\={{\equiv}}

\def\cross{\!\times\!}

\begin{document} 
\draft
\def\dbltopfraction{1.0}
\setlength{\dbltextfloatsep}{1pt plus 2pt minus 4pt}

\wideabs{
\title{\bf Symmetry of Magnetically Ordered Quasicrystals}
\author{Ron Lifshitz} 
\address{Condensed Matter Physics 114-36, California Institute of
Technology, Pasadena, CA 91125}
\maketitle

\begin{abstract}
The notion of magnetic symmetry is reexamined in light of the recent
observation of long range magnetic order in icosahedral quasicrystals
\protect{[}Charrier {\it et al.}, {\it Phys.\ Rev.\ Lett.} {\bf 78},
4637 (1997)\protect{]}.  The relation between the symmetry of a
magnetically-ordered (periodic or quasiperiodic) crystal, given in
terms of a ``spin space group,'' and its neutron diffraction diagram is
established. In doing so, an outline of a symmetry classification
scheme for magnetically ordered quasiperiodic crystals is
provided. Predictions are given for the expected diffraction patterns
of magnetically ordered icosahedral crystals, provided their
symmetry is well described by icosahedral spin space groups.
\end{abstract}
}

Quasicrystals, which today are known to exist in perfectly-ordered
thermodynamically-stable phases, have inspired a reexamination of the
basic notions of crystallinity, long-range order, and
symmetry~\cite{reviews}.  The recent observation by Charrier,
Ouladdiaf, and Schmitt~\cite{charrier} of long-range {\it magnetic
order\/} in icosahedral quasicrystals of composition
$R_8$Mg$_{42}$Zn$_{50}$ ($R$ = Tb, Dy, Ho, Er) is inspiring a similar
examination of the nature of magnetic order which, in quasicrystals,
had existed previously only as a theoretical construction~\cite{icq5}.
The purpose of this Letter is to provide the first steps in this
direction by explaining the notion of {\it magnetic symmetry\/} in a
manner which applies both to periodic and quasiperiodic crystals
thereby establishing a tool for future analysis of neutron
diffraction data and providing a tentative interpretation for the
findings of Charrier {\it et al.}~\cite{charrier}.

We choose to describe a magnetic material by its spin density field
$\sv(\rv)$. This field is a 3-component real-valued function,
transforming like an axial vector under $O(3)$ and changing sign under
time inversion. One may think of this function as
defining a set of classical magnetic moments, or spins, on the atomic
sites of the material. The standard expression (as shown, for example,
by Izyumov {\it et al.}~\cite{izy}) for the magnetic contribution to
the intensity of elastic scattering of unpolarized neutrons is then
given, in terms of the spin-density Fourier coefficients $\sv(\kv)$,
as
\begin{equation}\label{intensity}
I(\kv)\propto|\sv(\kv)|^2 - |\hat{\kv}\cdot\sv(\kv)|^2,
\end{equation} where $\kv$ is the scattering wave vector and
$\hat{\kv}$ is a unit vector in its direction.

In making this choice we follow Litvin and Opechowski~\cite{litvin}
who have developed a theory of ``spin space groups'' to describe the
symmetry of {\it periodic\/} magnetic crystals in terms of their spin
density fields. In their theory, symmetry operations are those leaving
the magnetic crystal {\it invariant.} These are the usual
3-dimensional space-group operations---translations and proper or
improper rotations---combined with rotations in ``spin space'' and
possibly also time inversion.  We extend their theory here to deal
with quasiperiodic crystals which possess neither translational
symmetry nor, in general, any rotations that leave them invariant.  We
choose to extend the theory of spin space groups rather than two other
commonly used theories for describing magnetic symmetry---those of
color symmetry~\cite{color} and the theory of representations of
ordinary space groups~\cite{bertaut}---because of its direct
predictions regarding the outcome of neutron scattering experiments.

What should we expect to see in a neutron diffraction diagram of a
single magnetically-ordered crystal? To answer this question we must
first clarify what we mean by ``crystal.'' The International Union of
Crystallography~\cite{iucr} defines a {\it crystal\/} to be ``\ldots
any solid with an essentially discrete diffraction diagram.''  To be
more concrete we consider spin density fields with well defined
Fourier transforms,
\begin{equation}\label{fourierS}
\sv(\rv)=\sum_{\kv\in L} \sv(\kv) e^{i\kv\cdot\rv},
\end{equation} in which the set $L$ contains at most a countable
infinity of plane waves.  In a real experiment, due to the finite
resolution of the apparatus, only a finite number of peaks whose
intensities $I(\kv)$ are above a certain threshold will be observed,
resulting in a discrete diffraction diagram.

What more can we say about the set of diffraction peaks beyond their
being essentially discrete?  We have shown elsewhere~\cite{krakow}
that if $\sv(\rv)$ describes a {\it physically stable\/} magnetic
crystal, \ie, one which minimizes a suitable Gibbs free energy, then
the wave vectors $\kv$ at which $\sv(\kv)\neq0$ are closed under
addition and subtraction, with the only exception of peaks that are
required by symmetry to vanish.  This implies in practice that once
enough peaks have been observed, additional peaks will appear with
increased experimental resolution only at integral linear combinations
of peaks that already exist.

This leads us to define the set $L$ in Eq.~\(fourierS) to be the set
of all integral linear combinations of the wave vectors $\kv$
determined by the diffraction diagram. We call this set the
(reciprocal) {\it magnetic lattice\/} of the crystal.  The {\it
rank\/} $D$ of $L$ is the smallest number of wave vectors needed to
generate it by taking integral linear combinations. As described
above, we expect that in most (if not all) experimentally observed
magnetic crystals this number will be finite. Periodic 3-dimensional
crystals have rank $D=3$; aperiodic 3-dimensional crystals have a rank
$D>3$; icosahedral quasicrystals, for example, have rank 6. The first
indication of the symmetry of the magnetic crystal is given by the set
of (proper or improper) rotations which, when applied to the origin of
Fourier space, merely permute the wave vectors of the magnetic
lattice. This set forms a subgroup of $O(3)$ called the {\it lattice
point group\/} $G_L$ (also called the holohedry).

The definition of ``lattice'' here is the same as in non-magnetic
crystals, except that the magnetic lattice supports the Fourier
transform of a vector function $\sv(\rv)$ whereas the lattice of a
non-magnetic crystal supports the Fourier transform of a scalar
function---that of the electronic or nuclear density $\rho(\rv)$ of
the crystal. We can therefore adopt the symmetry classification scheme,
used in the non-magnetic case, also for megnatic lattices. This
classification arranges lattices that have the same rank and the same
lattice point group $G_L$ into distinct {\it Bravais classes\/}
according to the way in which their vectors transform under $G_L$ (for
details see, for example, Dr\"ager and Mermin~\cite{jorg}).
Icosahedral lattices, for example, are arranged into three Bravais
classes~\cite{icosa}: $P$-lattices (primitive) contain all integral
linear combinations of the six vectors
\begin{equation}\label{vi}
\vv^{(1,4)}=(\pm1,\tau,0);\
\vv^{(2,5)}=(\tau,0,\pm1);\
\vv^{(3,6)}=(0,\pm1,\tau); 
\end{equation} where $\tau$ is the golden mean; $F^*$-lattices (face
centered in Fourier space) contain only those combinations in which
the sum of the six integers is even; and $I^*$-lattices (body centered
in Fourier space) contain only those combinations in which the six
integers are either all even or all odd.

To say anything further about the nature of the diffraction peaks we
must examine the symmetry of the spin density itself.  We reformulate
the theory of spin space groups by following the ideas of Rokhsar,
Wright, and Mermin's ``Fourier-space approach'' to
crystallography~\cite{rwm}. At the heart of this approach is a
redefinition of the concept of 3-dimensional point-group symmetry
which enables one to treat quasicrystals directly in 3-dimensional
space~\cite{volc}.  The key to redefining point-group symmetry is the
observation that certain rotations (proper or improper), when applied
to a quasiperiodic crystal, even though they do not leave the crystal
invariant, take it into one that contains the same spatial
distributions of bounded structures of arbitrary size. One finds that
any bounded region in the unrotated crystal is reproduced some
distance away in the rotated crystal, but there is, in general, no
single translation that brings the two crystals into perfect
coincidence.

This weaker notion of symmetry, termed ``indistinguishability,'' is
captured by requiring that any symmetry operation of the magnetic
crystal leave invariant all spatially-averaged autocorrelation
functions of its spin density field $\sv(\rv)$ for any order and for
any choice of components,
\begin{eqnarray}\label{corr2}
 \lefteqn{C^{(n)}_{\alpha_1\ldots\alpha_n}(\rv_1,\ldots,\rv_n)}
 \quad\quad \nonumber \\
 & & =\lim_{V\to\infty}{1\over V}\int_V d\rv 
   S_{\alpha_1}(\rv_1-\rv)\cdots S_{\alpha_n}(\rv_n-\rv).
\end{eqnarray} 

We have proven elsewhere (Appendix of~\cite{rmp}) that an equivalent
statement for the indistinguishability of any two quasiperiodic
multicomponent fields, in particular two spin density fields
$\sv(\rv)$ and $\sv'(\rv)$, is that their Fourier coefficients are
related by
\begin{equation}\label{chidef} 
\sv'(\kv) = e^{2\pi i\chi(\kv)}\sv(\kv), 
\end{equation} 
where $\chi$ is a real-valued linear function (modulo integers) on $L$
called a {\it gauge function.} Only in the case of periodic crystals
can one replace $2\pi \chi(\kv)$ by $\kv\cdot\dv$, reducing
indistinguishability to the requirement that the two crystals differ
at most by a translation $\dv$.

\begin{table*}[t]
\caption{Icosahedral lattice spin groups and their effect on the
outcome of neutron scattering experiments. Nontrivial lattice spin
groups are possible only with primitive ($P$) magnetic (reciprocal)
lattices or with body centered ($I^*$) magnetic (reciprocal)
lattices. The lattice spin groups, which are groups of rotations in
``spin space,'' are specified in the leftmost column in terms of their
generating rotations, where $2_i$ is a 2-fold rotation about the
$i$-axis in ``spin space'' ($i=x,y$, or $z$), $1'$ is the time
inversion operation which takes every $\sv(\kv)$ into $-\sv(\kv)$, and
${2'}_i$ is the product of the two. The form of $\sv(\kv)$ is given
for each scattering wave vector in the magnetic lattice according to
its indexing by the six vectors $\vv^{(i)}$ defined in Eq.~\(vi). In
the case of $I^*$-lattices, three different scattering patterns are
possible for each lattice spin group depending on the scale chosen for
the generating vectors $\vv^{(i)}$. Entries in brackets are related to
the ones above them through a scaling of the $I^*$-lattice by the
golden mean. The forms of the $\sv(\kv)$ given in the table may be
used in conjunction with expressions such as Eq.~\(intensity) to
determine the outcome of neutron scattering experiments.}
\label{table}
\begin{center}
\begin{footnotesize}
\setlength{\extrarowheight}{2pt}
\setlength{\tabcolsep}{8pt}
\begin{tabular}{|>{$}c<{$}|>{$}c<{$}>{$}c<{$}|>{$}c<{$}>{$}c<{$}>{$}c<{$}>{$}c<{$}|}
\cline{2-7}
\multicolumn{1}{c|}{} 
 &\multicolumn{2}{c|}{{\normalsize $P$ lattices---any integers $n_i$}}
 &\multicolumn{4}{c|}{{\normalsize $I^*$ lattices---$n_i$ all even or all odd}}\\
\cline{2-7}
\multicolumn{1}{c|}{}
&& & {\underbar{$n_i$ all even}} & {\underbar{$n_i$ all even}} 
   & {\underbar{$n_i$ all odd}} & {\underbar{$n_i$ all odd}}\\
\multicolumn{1}{c|}{}
   & {\sum n_i = 2n} & {\sum n_i = 2n+1} 
   & {\sum n_i = 4n} & {\sum n_i = 4n+2} 
   & {\sum n_i = 4n} & {\sum n_i = 4n+2}\\ 
\hline
&& & (0,0,0) & (0,0,0) & (S_x,S_y,S_z) & (S_x,S_y,S_z) \\
1' & (0,0,0) & (S_x, S_y, S_z) 
   & [(0,0,0)] & [(S_x,S_y,S_z)] & [(S_x,S_y,S_z)] & [(0,0,0)] \\
&& & [(0,0,0)] & [(S_x,S_y,S_z)] & [(0,0,0)] & [(S_x,S_y,S_z)]  \\
\hline
&&  &(0,0,S_z) &(0,0,S_z) & (S_x,S_y,0) & (S_x,S_y,0) \\
2_z & (0, 0, S_z) & (S_x, S_y, 0) 
    & [(0,0,S_z)] & [(S_x,S_y,0)] & [(S_x,S_y,0)] & [(0,0,S_z)]  \\
&&  & [(0,0,S_z)] & [(S_x,S_y,0)] & [(0,0,S_z)] & [(S_x,S_y,0)]  \\
\hline
&&     & (S_x,S_y,0) & (S_x,S_y,0) &(0,0,S_z) &(0,0,S_z) \\
{2'}_z & (S_x, S_y, 0) & (0, 0, S_z)
       & [(S_x,S_y,0)] & [(0,0,S_z)] & [(0,0,S_z)] & [(S_x,S_y,0)]  \\
&&     & [(S_x,S_y,0)] & [(0,0,S_z)] & [(S_x,S_y,0)] & [(0,0,S_z)]  \\
\hline
&&  & (0,0,0) & (0,S_y,0) & (S_x,0,0) & (0,0,S_z) \\
2_x2_y2_z & \multicolumn{2}{c|}{N/A}
    & [(0,0,0)] & [(S_x,0,0)] & [(0,0,S_z)] & [(0,S_y,0)] \\
&&  & [(0,0,0)] & [(0,0,S_z)] & [(0,S_y,0)] & [(S_x,0,0)]  \\
\hline
&&   & (0,0,S_z) & (S_x,0,0) & (0,S_y,0) & (0,0,0) \\
{2'}_x{2'}_y2_z & \multicolumn{2}{c|}{N/A}
   & [(0,0,S_z)] & [(0,S_y,0)] & [(0,0,0)] & [(S_x,0,0)] \\
&& & [(0,0,S_z)] & [(0,0,0)] & [(S_x,0,0)] & [(0,S_y,0)] \\
\hline
&& & (0,0,0) & (0,0,0) &(0,0,S_z) & (S_x,S_y,0) \\
2_z1' & \multicolumn{2}{c|}{N/A}
   & [(0,0,0)] & [(0,0,S_z)] & [(S_x,S_y,0)] & [(0,0,0)] \\
&& & [(0,0,0)] & [(S_x,S_y,0)] & [(0,0,0)] & [(0,0,S_z)] \\
\hline
\end{tabular}
\end{footnotesize}
\end{center}
\end{table*}

We can now define the {\it point group $G$\/} of the magnetic crystal
to be the set of operations $g$ from $O(3)$ that leave it
indistinguishable to within rotations $\g$ in spin space, possibly
combined with time inversion. Accordingly, for every pair $(g,\g)$
there exists a gauge function, $\Phi_g^\g(\kv)$, called a {\it phase
function}, which satisfies
\begin{equation}\label{phase}
\sv(g\kv) = e^{2\pi i \Phi_g^\g(\kv)}\g\sv(\kv).
\end{equation} Since $\sv([gh]\kv)=\sv(g[h\kv])$, one easily
establishes that the transformations $\g$ in spin space form a group
$\G$ and that the pairs $(g,\g)$ satisfying the point-group condition
\(phase) form a subgroup of $G\times\G$ which we call the {\it spin
point group $\GS$}. The corresponding phase functions, one for each
pair in $\GS$, must satisfy the {\it group compatibility condition,}
\begin{equation}\label{GCC}
\forall (g,\g), (h,\eta)\in\GS:\quad \Phi_{gh}^{\g\eta}(\kv) \=
\Phi_g^\g(h\kv) + \Phi_h^\eta(\kv),
\end{equation} where ``$\=$'' denotes equality modulo integers. A
{\it spin space group,} describing the symmetry of a magnetic crystal,
whether periodic or aperiodic, is thus given by a magnetic lattice
$L$, a spin point group $\GS$, and a set of phase functions
$\Phi_g^\g(\kv)$, satisfying the group compatibility condition \(GCC).

In order to identify further the common symmetry properties of
different magnetic structures, whose lattices and spin point groups
are equivalent, one classifies their spin space groups into properly
chosen equivalence classes called {\it spin space-group types.} This
is achieved by organizing sets of phase functions satisfying the group
compatibility condition \(GCC) into equivalence classes. Two such sets
$\Phi$ and $\Phi'$ are equivalent if: (I) they describe
indistinguishable spin density fields, related as in Eq.~\(chidef) by
a gauge function $\chi$; or (II) they correspond to alternative
descriptions of the same crystal that differ by their choices of
absolute length scales and spatial orientations. In case (I) $\Phi$
and $\Phi'$ are related by a {\it gauge transformation}:
\begin{equation}\label{gauge-tr}
\forall (g,\g)\in\GS:\quad {\Phi'}_{g}^{\g}(\kv) \= 
\Phi_{g}^{\g}(\kv) + \chi(g\kv-\kv).
\end{equation} For a more rigorous definition of these equivalence
criteria see the analogous classification of color space
groups (Sec.~III of~\cite{rmp}).

We said earlier that every wave vector $\kv$ in the magnetic lattice
$L$ of a magnetic crystal is a candidate for a diffraction peak unless
symmetry forbids it. We are now in a position to understand how this
happens.  Given a wave vector $\kv\in L$ we examine all spin
point-group operations $(g,\g)$ for which $g\kv=\kv$.  These elements
form a subgroup of the spin point group which we call the {\it little
spin group of $\kv$}, $\GSk$. For elements $(g,\g)$ of $\GSk$, the
point-group condition \(phase) can be rewritten as
\begin{equation}\label{eigen}
\g\sv(\kv) = e^{-2\pi i \Phi_g^\g(\kv)}\sv(\kv).
\end{equation} This implies that every Fourier coefficient $\sv(\kv)$
is required to be a simultaneous eigenvector of all spin
transformations $\g$ in the little spin group of $\kv$, with the
eigenvalues given by the corresponding phase functions. If a
non-trivial 3-dimensional vector satisfying Eq.~\(eigen) does not
exist then $\sv(\kv)$ will necessarily vanish. It should be noted that
the phase values in Eq.~\(eigen), are independent of the choice of
gauge~\(gauge-tr), and are therefore uniquely determined by the spin
space-group type of the crystal.

The process of determining the form of the simultaneous eigenvector
$\sv(\kv)$ is greatly simplified if one makes the following
observation.  Due to the group compatibility condition \(GCC) the set
of eigenvalues in Eq.~\(eigen) for all the elements $(g,\g)\in\GSk$
form a 1-dimensional representation of that group. Spin space-group
symmetry thus requires the Fourier coefficient $\sv(\kv)$ to transform
under a particular 1-dimensional representation of the spin
transformations in the little spin group of $\kv$. We also
independently know that $\sv(\kv)$ transforms under spin rotations as
a 3-dimensional axial vector, changing its sign under time
inversion. We therefore need to check whether the particular
1-dimensional representation, dictated by the spin space group, is
contained within the 3-dimensional axial-vector representation. If it
is not, then $\sv(\kv)$ must vanish; if it is, then $\sv(\kv)$ must
lie in the subspace of spin space transforming under this
1-dimensional representation.

Of particular interest are spin transformations $\g$ that leave the
spin density field indistinguishable without requiring any rotation in
physical space. These transformations are paired in the spin point
group with the identity rotation $e$ and form a subgroup of $\G$
called the {\it lattice spin group $\G_e$.}  The lattice spin group
plays a key role in determining the outcome of elastic neutron
scattering, for if a magnetic crystal has a nontrivial lattice spin
group $\G_e$ then $\{e\}\cross\G_e \subseteq\GS^\kv$ for every $\kv$
in the magnetic lattice, restricting the form of the corresponding
$\sv(\kv)$.  In the special case of periodic crystals, the elements
of $\G_e$ are spin transformations that when combined with
translations leave the magnetic crystal invariant. The phase functions
$\Phi_e^\g(\kv)$ therefore contain the information which generalizes
to the quasiperiodic case the so-called ``spin translation groups'' of
Litvin and Opechowski~\cite{litvin}.

What are the possible values of the phase functions $\Phi_e^\g(\kv)$?
Consider, for example, a lattice spin group generated by an $n$-fold
rotation $\g$ about the $z$-axis in spin space, with $n>2$. Repeated
applications of the group compatibility condition \(GCC) to
$(e,\g)^n=(e,\eps)$, where $\eps$ is the identity in spin space, give
$0\= \Phi_e^{\g^n}(\kv) \= n\Phi_e^\g(\kv)$. Thus $\Phi_e^\g(\kv) \=
j/n$ for some integer $j$. One can then easily verify through
Eq.~\(eigen) that
\begin{equation}\label{example2}
\sv(\kv) = \cases{ (0,0,S_z) & $\Phi_e^\g(\kv)\=0$,\cr 
(S_\perp,\pm iS_\perp,0) &
$\Phi_e^\g(\kv)\=\pm\frac1{n}$,\cr (0,0,0) & otherwise.\cr
}\end{equation}

We have enumerated the distinct lattice spin groups for icosahedral
quasicrystals and have found that non-trivial lattice spin groups are
possible only with $P$- or $I^*$-magnetic lattices. For each of the
lattice spin groups we have also determined the expected form of
$\sv(\kv)$, that is required through \(eigen) by the spin space group
symmetry, for every scattering wave vector $\kv$ in the magnetic
lattice. These results are summarized in Table~\ref{table}.  Further
restrictions (not tabulated here) may exist for wave vectors $\kv$
lying in the invariant subspaces of non-trivial point-group
operations. We emphasize that the diffraction patterns are described
here as magnetic lattices with missing points rather than nuclear
lattices that are shifted by so-called ``magnetic propagation
vectors.''

Charrier {\it et al.}~\cite{charrier} observe magnetic reflections for
the $R_8$Mg$_{42}$Zn$_{50}$ quasicrystals at wave vectors of the form
$\kv = \sum_{i=1}^6 m_i \vv^{(i)} +p \vv^{(j)}$, where the $\vv^{(i)}$
are defined in Eq.~\(vi), the $m_i$ are either all even or all odd,
$j=1,\ldots, 6$, and $p=\pm\frac12$. The nuclear reflections form a
body-centered icosahedral (reciprocal) lattice, obtained from the
expression above but with $p=0$, corresponding to face-centered
ordering in direct space. If the magnetic structure indeed has
icosahedral symmetry, and is not merely a collection of magnetic
domains in which icosahedral symmetry is broken, then the magnetic
lattice, which is formed by taking all integral linear combinations of
the observed magnetic reflections, is a primitive icosahedral lattice
containing all vectors of the form $\kv = \sum_{i=1}^6 n_i (\frac12
\vv^{(i)})$ with any integers $n_i$. All reflections at wave vectors
with $\sum_{i=1}^6 n_i$ even are not observed by Charrier {\it et al.}
which is consistent with having a lattice spin group $1'$, as shown by
the first entry for $P$-lattices in Table~\ref{table}. Other peaks
that are not observed might just be too weak in the current experiment
rather than actually missing due to symmetry requirements. This may
suggest that in direct space the magnetic structure has an underlying
antiferromagnetic body-centered icosahedral ordering (analogous to the
chemical ordering in the cubic cesium chloride structure). This may
occur, for example, if only a fraction of the rare-earth atoms have
their magnetic moments aligned, this fraction arranged in a
body-centered icosahedral super-structure made of tiles with twice the
edge length. One would need to develop actual models to test such
hypotheses.

We have demonstrated that through its fairly simple selection rules
\(eigen) the theory of spin space groups provides a valuable tool for
analyzing neutron diffraction diagrams of either periodic or
quasiperiodic magnetically-ordered crystals. This is not to say that
the use of color symmetry, which has been extended to
quasicrystals~\cite{rmp}, and the use of representations of ordinary
space groups, which have yet to be extended to quasicrystals, will not
offer any additional insight.

This paper is dedicated to Hans-Ude Nissen on the occasion of his
65{\it th\/} birthday.  The author thanks Michael Cross, Veit Elser,
Peter Weichman, Beno\^{\i}t Charrier, and Denys Schmitt for valuable
discussions, and the latter two also for sending their preprints prior
to publication.  This work was supported by Caltech through a Division
Research Fellowship in Theoretical Physics.

\end{document}